\documentclass{llncs}
\usepackage{llncsdoc}
\usepackage{graphicx}
\usepackage{caption}
\usepackage{subcaption}
\begin{document}

\title{Digital Stylometry: Linking Profiles Across Social Networks}

\author{Soroush Vosoughi \and Helen Zhou \and Deb Roy}
\institute{MIT Media Lab, Cambridge, MA, USA \\ \email{soroush@mit.edu, hlzhou@mit.edu, dkroy@media.mit.edu}}

\maketitle

\begin{abstract}
There is an ever growing number of users with accounts on multiple social media and networking sites. Consequently, there is increasing interest in matching user accounts and profiles across different social networks in order to create aggregate profiles of users. In this paper, we present models for \emph{Digital Stylometry}, which is a method for matching users through \emph{stylometry} inspired techniques. We experimented with linguistic, temporal, and combined temporal-linguistic models for matching user accounts, using standard and novel techniques. Using publicly available data, our best model, a combined temporal-linguistic one, was able to correctly match the accounts of $31\%$ of $5,612$ distinct users across Twitter and Facebook. 

\end{abstract}

%
\section{Introduction}
Stylometry is defined as, "the statistical analysis of variations in literary style between one writer or genre and another". It is a centuries-old practice, dating back the early Renaissance. It is most often used to attribute authorship to disputed or anonymous documents. Stylometry techniques have also successfully been applied to other, non-linguistic fields, such as paintings and music. The main principles of stylometry were compiled and laid out by the philosopher Wincenty Lutosławski in 1890 in his work "Principes de stylométrie" \cite{styleone}.

Today, there are millions of users with accounts and profiles on many different social media and networking sites. It is not uncommon for users to have multiple accounts on different social media and networking sites. With so many networking, emailing, and photo sharing sites on the Web, a user often accumulates an abundance of account profiles. There is an increasing focus from the academic and business worlds on aggregating user information across different sites, allowing for the development of more complete user profiles. There currently exist several businesses that focus on this task \cite{b1,b2,b3}. These businesses use the aggregate profiles for advertising, background checks or customer service related tasks. Moreover, profile matching across social networks, can assist the growing field of social media rumor detection \cite{qazvinian2011rumor,takahashi2012rumor,vosoughi2015automatic,vosoughihuman},   since many malicious rumors are spread on different social media platforms by the same people, using different accounts and usernames.

Motivated by traditional stylometry and the growing interest in matching user accounts across Internet services, we created models for \emph{Digital Stylometry}, which fuses traditional stylometry techniques with big-data driven social informatics methods used commonly in analyzing social networks. Our models use linguistic and temporal activity patterns of users on different accounts to match accounts belonging to the same person. We evaluated our models on $11,224$ accounts belonging to $5,612$ distinct users on two of the largest social media networks, Twitter and Facebook. The only information that was used in our models were the time and the linguistic content of posts by the users. We intentionally did not use any other information, especially the potentially personally identifiable information that was explicitly provided by the user, such as the screen name, birthday or location. This is in accordance with traditional stylometry techniques, since people could misstate, omit, or lie about this information. Also, we wanted to show that there are implicit clues about the identities of users in the content (language) and context (time) of the users' interactions with social networks that can be used to link their accounts across different services.

Other than the obvious technical goal, the purpose of this paper is to shed light on the relative ease with which seemingly innocuous information can be used to track users across social networks, even when signing up on different services using completely different account and profile information (such as name and birthday). This paper is as much of a technical contribution, as it is a warning to users who increasingly share a large part of their private lives on these services.

The rest of this paper is structured as follows. In the next sections we will review related work on linking profiles, followed by a description of our data collection and annotation efforts. After that, we discuss the linguistic, temporal and combined temporal-linguistic models developed for linking user profiles.  Finally, we discuss and summarize our findings and contributions and discuss possible paths for future work. 

\section{Related Work}
There are several recent works that attempt to match profiles across different Internet services. Some of these works utilize private user data, while some, like ours, use publicly available data. An example of a work that uses private data is Balduzzi et al. \cite{balduzzi2010abusing}. They use data from the \emph{Friend Finder} system (which includes some private data) provided by various social networks to link users across services. Though one can achieve a relatively high level of success by using private data to link user accounts, we are interested in using only publicly available data for this task. In fact, as mentioned earlier, we do not even consider publicly available information that could explicitly identify a user, such as names, birthdays and locations.

Several methods have been proposed for matching user profiles using public data \cite{20,34,33,29,27,35,labitzke2011your,14}. These works differ from ours in two main aspects. First, in some of these works, the ground truth data is collected by assuming that all profiles that have the same screen name are from the same users \cite{labitzke2011your,14}. This is not a valid assumption. In fact, it has been suggested that close to $20\%$ of accounts with the same screen name in Twitter and Facebook are not matching \cite{24}. Second, almost all of these works use features extracted from the user profiles \cite{20,34,33,29,27,35,labitzke2011your}. Our work, on the other hand, is blind to the profile information and only utilizes users' activity patterns (linguistic and temporal) to match their accounts across different social networks. Using profile information to match accounts is contrary to the best practices of stylometry since it assumes and relies on the honesty, consistency and willingness of the users to explicitly share identifiable information about themselves (such as location).

\section{Data Collection and Datasets}
For the purposes of this paper, we focused on matching accounts between two of the largest social networks: Twitter and Facebook. In order to proceed with our study, we needed a sizeable (few thousand) number of English speaking users with accounts on both Twitter and Facebook. We also needed to know the precise matching between the Twitter and Facebook accounts for our ground truth.


To that end, we crawled publicly available, English-language, Google Plus accounts using the Google Plus API \footnote{https://developers.google.com/+/web/api/rest/} and scraped links to the users' other social media profiles. (Note that one of the reasons why we used Twitter and Facebook is that they were two of the most common sites linked to on Google Plus). We used a third party social media site (i.e., Google Plus), one that was not used in our analysis to compile our ground truth in order to limit selection bias in our data collection. 

We discarded all users who did not link to an account for both Twitter and Facebook and those whose accounts on either of these sites were not public. We then used the APIs of Twitter\footnote{https://dev.twitter.com/rest/public} and Facebook\footnote{https://developers.facebook.com/docs/public\_feed} to collect posts made by the users on these sites. We only collected  the linguistic content and the date and time at the which the posts were made. For technical and privacy reasons, we did not collect any information from the profile of the users, such as the location, screen name, or birthday. 

Our analysis focused on activity of users for one whole year, from February 1st, 2014 to February 1st, 2015. Since we can not reliably model the behaviour patterns of users with scarce data, users with less than $20$ posts in that time period on either site were discarded. Overall, we collected a dataset of $5,612$ users with each having a Facebook and Twitter account for a total of $11,224$ accounts. 

\begin{figure}[h]
\centering
\begin{subfigure}{0.49\textwidth}
\includegraphics[width=\linewidth]{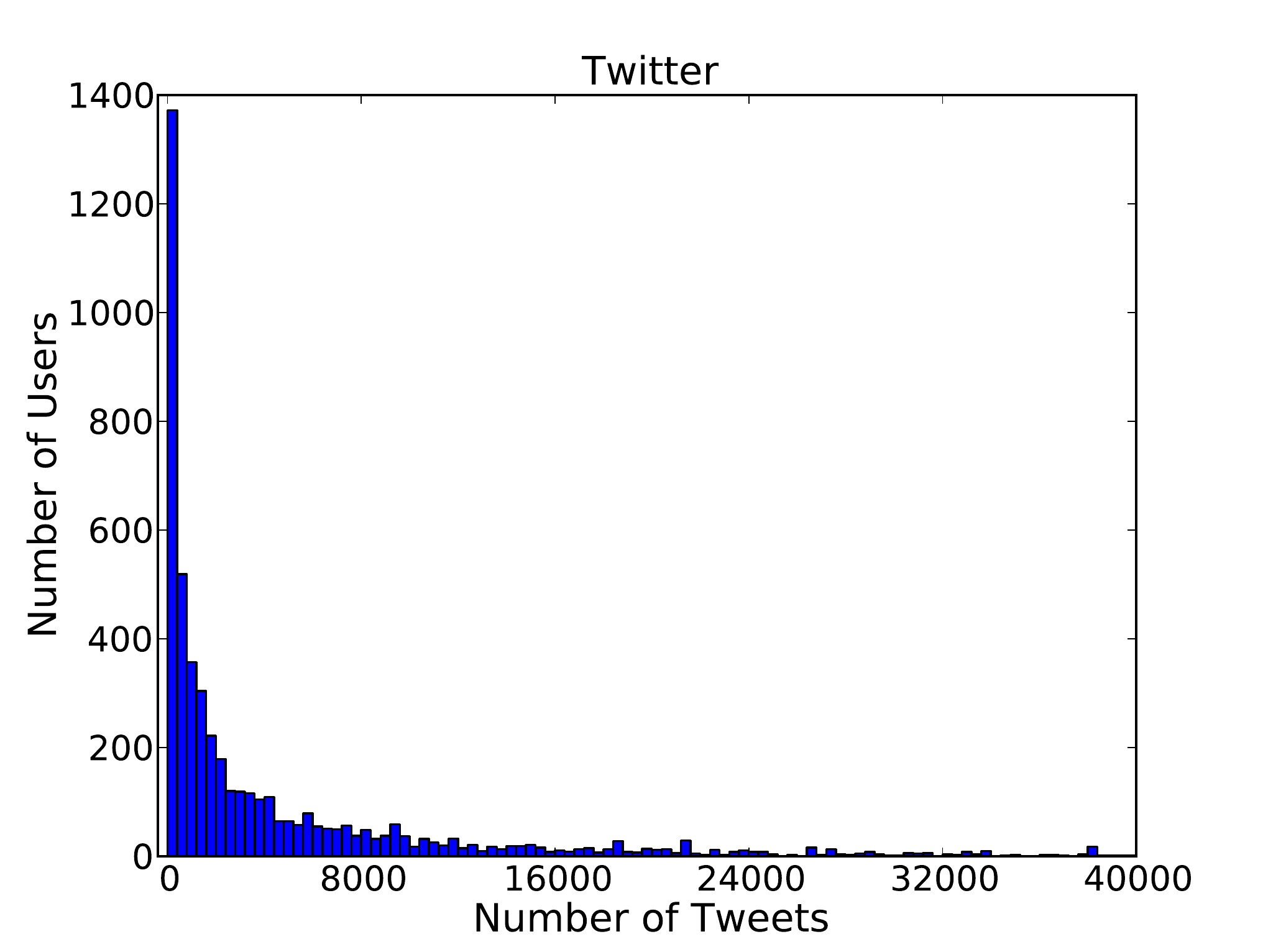}
\caption{Twitter data.}
\label{fig:tdata}
\end{subfigure}
\begin{subfigure}{0.49\textwidth}
\includegraphics[width=\linewidth]{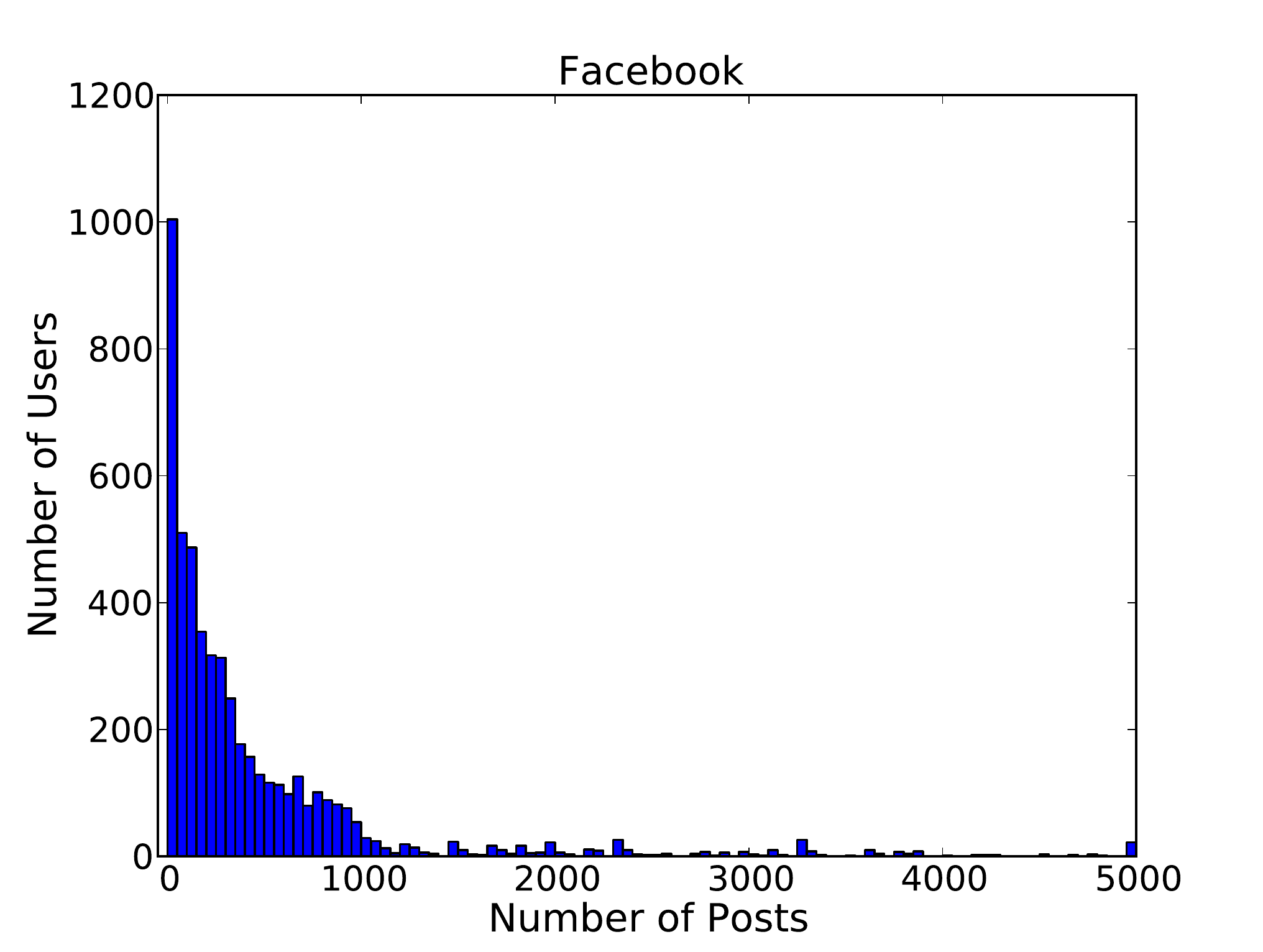}
\caption{Facebook data.}
\label{fig:fdata}
\end{subfigure}
\caption{Distribution of number of posts per user for Twitter and Facebook, from our collected dataset.}
\label{fig:data}
\end{figure}

Figure \ref{fig:data} shows the distribution of the number of posts per user for Twitter and Facebook for our collected dataset. In the figure, the data for the number of posts has been divided into $500$ bins. For the Twitter data, each bin corresponds to $80$ tweets, while for the Facebook data, it corresponds to $10$ posts. Table \ref{tab:data} shows some statistics about the data collected, including the average number of posts per user for each of the sites.

\begin{table}
\centering
\caption{Statistics about the number of posts by the users of the $5,612$ accounts collected from Twitter and Facebook.}

\begin{tabular}{l|ll}	
\hline			
	 &{Twitter} 		& {Facebook}			\\ \hline
Mean & 1,535 & 155 \\
Median & 352   & 54 	\\
Maximum		& 39,891 	& 4,907		\\
Minimum	& 20	 &	20  \\
\hline
\end{tabular}
\label{tab:data}
\end{table}

\section{Models}
We developed several linguistic, temporal and combined temporal-linguistic models for our task. These models take as input a user, $u$, from one of the sites (i.e., Twitter or Facebook) and a list of $N$ users from the other service, where one of the $N$ users, $u\prime$, is the same as $u$. The models then provide a ranking among candidate matches between $u$ and each of the $N$ users. We used two criteria to evaluate our models: 
\begin{itemize}
\item{Accuracy: percentage of cases when a model's top ranked candidate is $u\prime$.}
\item{Average Rank: the average rank of $u\prime$ within the ranked list of candidates generated by a model.}
\end{itemize}

A baseline random choice ranker would have an accuracy of $1/N$, and an average rank of $N/2$ (since $u\prime$ may appear anywhere in the list of $N$ items). 

\subsection{Linguistic Models}
A valuable source of information in matching user accounts, one used in traditional stylometry tasks, is the way in which people use language. A speaker or writer's choice of words depends on many factors, including the rules of grammar, message content and stylistic considerations. There is a great variety of possible ways to compare the language patterns of two people. However, first we need a method for modelling the language of a given user. Below we explain how this is done.

\subsubsection{Language Models}
Most statistical language models do not attempt to explicitly model the complete language generation process, but rather seek a compact model that adequately explains the observed linguistic data. Probabilistic models of language assign probabilities to word sequences $w_1$ . . . $w_\ell$, and as such the likelihood of a corpus can be used to fit model parameters as well as characterize model performance.

N-gram language modelling \cite{eugene1996statistical,jurafsky2000speech,manning1999foundations} is an effective technique that treats words as samples drawn from a distribution conditioned on other words, usually the immediately preceding $n-1$ words, in order to capture strong local word dependencies. The probability of a sequence of $\ell$ words, written compactly as $w_1^\ell$ is $\Pr(w_1^\ell)$ and can be factored exactly as 
$$\Pr(w_1^\ell) = \Pr(w_1) \prod_{i=2}^\ell \Pr(w_i|w_1^{i-1})$$ 

However, parameter estimation in this full model is intractable, as the number of possible word combinations grows exponentially with sequence length. N-gram models address this with the approximation $\tilde{\Pr}(w_i|w_{i-n+1}^{i-1}) \approx \Pr(w_i|w_1^{i-1})$ using only the preceding $n-1$ words for context. A bigram model ($n=2$) uses the preceding word for context, while a unigram model ($n=1$) does not use any context.


For this work, we used unigram models in Python, utilizing some components from NLTK \cite{bird2009natural}. Probability distributions were calculated using Witten-Bell smoothing \cite{jurafsky2000speech}. Rather than assigning word $w_i$ the maximum likelihood probability estimate $p_i = \frac{c_i}{N}$, where $c_i$ is the number of observations of word $w_i$ and $N$ is the total number of observed tokens, Witten-Bell smoothing discounts the probability of observed words to $p_i^* = \frac{c_i}{N+T}$ where $T$ is the total number of observed word types. The remaining $Z$ words in the vocabulary that are unobserved (i.e. where $c_i = 0$) are given by $p_i^* = \frac{T}{Z(N+T)}$. 

We experimented with two methods for measuring the similarity between n-gram language models. In particular, we tried approaches based on \emph{KL-divergence} and \emph{perplexity} \cite{cover2012elements}. We also tried two methods that do not rely on n-gram models, \emph{cosine similarity of TF-IDF vectors} \cite{rajaraman2011mining}, as well as our own novel method, called the \emph{confusion model}. 

The performance of each method is shown in Table \ref{table:ling_results}. Note that all methods outperform the random baseline in both accuracy and average rank by a great margin. Below we explain each of these metrics.

\begin{table}

\centering
\caption{Performance of different linguistic models, tested on 5,612 users (11,224 accounts), sorted by accuracy. Best results are shown bold.}

\begin{tabular}{l|ll}
\hline
				& \multicolumn{2}{c}{Performance} \\
{Model}	 &{ Accuracy} 		& {AverageRank}			\\ \hline
Random Baseline & 0.0002 & 2,806\\
Perplexity & 0.06   & 1,498 	\\
KL-divergence		& 0.08 	& 2,029		\\
TF-IDF	& 0.21	&	999  \\
Confusion	& \bf{0.27} 	&	\bf{859} \\
\hline
\end{tabular}
\label{table:ling_results}
\end{table}


\subsubsection{KL-divergence}
The first metric used for measuring the distance between the language of two user accounts is the Kullback-Leibler (KL) divergence \cite{cover2012elements} between the unigram probability distribution of the corpus corresponding to the two accounts. The KL-divergence provides an asymmetric measure of dissimilarity between two probability distribution functions $p$ and $q$ and is given by:

$$KL(p||q) = \int p(x)ln\frac{p(x)}{q(x)}$$ 

We can modify the equation to prove a symmetric distance between distributions:

$$KL_{2}(p||q) = KL(p||q)+KL(q||p)$$

\subsubsection{Perplexity}
For this method, the similarity metric is the perplexity \cite{cover2012elements} of the unigram language model generated from one account, $p$ and evaluated on another account, $q$. Perplexity is given as:

$$PP(p,q) = 2^{H(p,q)}$$ 

where $H(p,q)$ is the cross-entropy \cite{cover2012elements} between distributions of the two accounts $p$ and $q$. More similar models lead to smaller perplexity. As with KL-divergence, we can make perplexity symmetric:

$$PP_{2}(p,q) = PP(p,q)+PP(q,p)$$


This method outperformed the \emph{KL-divergence} method in terms of average rank but not accuracy (see Table \ref{table:ling_results}).

\subsubsection{TF-IDF}
Perhaps the relatively low accuracies of perplexity and KL-divergence measures should not be too surprising. These measures are most sensitive to the variations in frequencies of most common words. For instance, in its most straightforward implementation, the KL-divergence measure would be highly sensitive to the frequency of the word ``the". Although this problem might be mitigated by the removal of stop words and applying topic modelling to the texts, we believe that this issue is more nuanced than that. 

Different social media (such as Twitter and Facebook) are used by people for different purposes, and thus Twitter and Facebook entries by the same person are likely to be thematically different. So it is likely that straightforward comparison of language models would be inefficient for this task.

One possible solution for this problem is to look at users' language models not in isolation, but in comparison to the languages models of everyone else. In other words, identify features of a particular language model that are characteristic to its corresponding user, and then use these features to estimate similarity between different accounts. This is a task that \emph{Term Frequency-Inverse Document Frequency}, or TF-IDF, combined with \emph{cosine similarity}, can manage. 

TF-IDF is a method of converting text into numbers so that it can be represented meaningfully by a vector \cite{rajaraman2011mining}. TF-IDF is the product of two statistics, \emph{TF} or Term Frequency and \emph{IDF} or Inverse Document Frequency. Term Frequency measures the number of times a term (word) occurs in a document. Since each document will be of different size, we need to normalize the document based on its size. We do this by dividing the Term Frequency by the total number of terms. 

TF considers all terms as equally important, however, certain terms that occur too frequently should have little effect (for example, the term \emph{``the"}). And conversely, terms that occur less in a document can be more relevant. Therefore, in order to weigh down the effects of the terms that occur too frequently and weigh up the effects of less frequently occurring terms, an Inverse Document Frequency factor is incorporated which diminishes the weight of terms that occur very frequently in the document set and increases the weight of terms that occur rarely. Generally speaking, the Inverse Document Frequency is a measure of how much information a word provides, that is, whether the term is common or rare across all documents. 








Using TF-IDF, we derive a vector from the corpus of each account. We measure the similarity between two accounts using cosine similarity:

$$Similarity(d1,d2) = \frac{d1 \cdot d2}{||d1||\times ||d2||}$$

Here, \emph{$d1 \cdot d2$} is the dot product of two documents, and \emph{$||d1||\times ||d2||$} is the product of the magnitude of the two documents. Using TD-IDF and cosine similarity, we achieved significantly better results than the last two methods, with an accuracy of $0.21$ and average rank of $999$.

\subsubsection{Confusion Model} 
TF-IDF can be thought of as a heuristic measure of the extent to which different words are characteristic of a user. We came up with a new, theoretically motivated measure of ``being characteristic" for words. We considered the following setup :

\begin{enumerate}
\item The whole corpus of the $11,224$ Twitter and Facebook accounts was treated as one long string;
\item For each token in the string, we know the user who produced it. Imagine that we removed this information and are now making a guess as to who the user was. This will give us a probability distribution over all users;
\item Now imagine that we are making a number of the following samples: randomly selecting a word from the string, taking the \emph{true user}, $TU$ for this word and a guessed user, $GU$ from correspondent probability distribution. Intuitively, the more often a particular pair, $TU=U_{1}, GU=U_{2}$ appear together, the stronger is the similarity between $U_{1}$ and $U_{2}$;
\item We then use mutual information to measure the strength of association. In this case, it will be the mutual information \cite{cover2012elements} between random variables, $TU=U_{1}$ and $GU=U_{2}$. This mutual information turns out to be proportional to the probabilities of $U_{1}$ and $U_{2}$ in the dataset, which is undesirable for a similarity measure. To correct for this, we divide it by the probabilities of $U_{1}$ and $U_{2}$;  
\end{enumerate}

We call this model the \emph{confusion model}, as it evaluated the probability that $U_{1}$ will be confused for $U_{2}$ on the basis of a single word. The expression for the similarity value according to the model is $S\times log(S)$, where $S$ is:

$$S=\sum_{w} p(w)p(U_{1}|w)p(U_{2}|w)$$ 

Note that if $U_{1}=U_{2}$, the words contributing most to the sum will be ordered by their ``degree of being characteristic". The values, $p(w)$ and $p(u|w)$ have to be estimated from the corpus. To do that, we assumed that the corpus was produced using the following auxiliary model:
\begin{enumerate}
\item For each token, a user is selected from a set of users by multinomial distribution;
\item A word is selected from a multinomial distribution of words for this user to produce the token.
\end{enumerate}

We used Dirichlet distributions \cite{balakrishnan2004primer} as priors over multinomials. This method outperforms all other methods with an accuracy of $0.27$ and average rank of $859$.

\subsection{Temporal Models}
Another valuable source of information in matching user accounts, is the activity patterns of users. A measure of activity is the time and the intensity at which users utilize a social network or media site. All public social networks, including publicly available Twitter and Facebook data, make this information available. Previous research has shown temporal information (and other contextual information, such as spatial information) to be correlated with the linguistic activities of people \cite{roy2014grounding,vosoughi-zhou-roy:2015:WASSA}.

We extracted the following discrete temporal features from our corpus: month ($12$ bins), day of month ($31$ bins), day of week ($7$ bins) and hour ($24$ bins). We chose these features to capture fine to coarse-level temporal patterns of user activity. For example, commuting to work is a recurring pattern linked to a time of day, while paying bills is more closely tied to the day of the month, and vacations are more closely tied to the month.

We treated each of these bins as a word, so that we could use the same methods used in the last section to measure the similarity between the temporal activity patterns of pairs of accounts (this will also help greatly for creating the combined model, explained in the next section). In other word, the $12$ bins in month were set to $w_1$ . . . $w_{12}$, the $31$ bins in day of month to $w_{13}$ . . . $w_{43}$, the $7$ bins in day of week to $w_{44}$ . . . $w_{50}$, and the $24$ bins in time were set to $w_{51}$ . . . $w_{74}$. Thus, we had a corpus of $74$ words. 

For example, a post on \emph{Friday, August 5th at 2 AM} would be translated to $\{w_8,w_{17},w_{48},w_{53}\}$, corresponding to {August, 5th, Friday, 2 AM} respectively. Since we are only using unigram models, the order of words does not matter. As with the language models described in the last section, all of the probability distributions were calculated using Witten-Bell smoothing. We used the same four methods as in the last section to create our temporal models. 

Table \ref{table:temporal_results} shows the performance of each of these models. Although the performance of the temporal models were not as strong as the linguistic ones, they all vastly outperformed the baseline. Also, note that here as with the linguistic models, the \emph{confusion model} greatly outperformed the other models.


\begin{table}
\centering
\caption{Performance of different temporal models, tested on 5,612 users (11,224 accounts), sorted by accuracy. Best results are shown bold.}

\begin{tabular}{l|ll}
\hline
				& \multicolumn{2}{c}{Performance} \\
{Model}	 &{Accuracy} 		& {AverageRank}			\\ \hline
Random Baseline & 0.0002 & 2,806\\
KL-divergence		& 0.02 	& 2,491		\\
Perplexity & 0.03   & 2,083 	\\
TF-IDF	& 0.07	&	1,503  \\
Confusion	& \bf{0.10} 	&	\bf{1,458} \\
\hline
\end{tabular}
\label{table:temporal_results}
\end{table}


\subsection{Combined Models}
Finally, we created a combined temporal-linguistic model. Since both the linguistic and the temporal models were built using the same framework, it was fairly simple to combine the two models. The combined model was created by merging the linguistic and temporal corpora and vocabularies. (Recall that we treated temporal features as words). We then experimented with the same four methods as in the last two sections to create our combined models.

Table \ref{table:comb_results} shows the performance of each of these models. Across the board, the combined models outperformed their corresponding linguistic and temporal models, though the difference with the linguistic models were not as great. These results suggest that at some level the temporal and the linguistic "styles" of users provide non-overlapping cues about the identity of said users. Also, note that as with the linguistic and temporal models, our combined \emph{confusion model} outperformed the other combined models.

\begin{figure}[!htpb]
\centering
\includegraphics[width=\columnwidth]{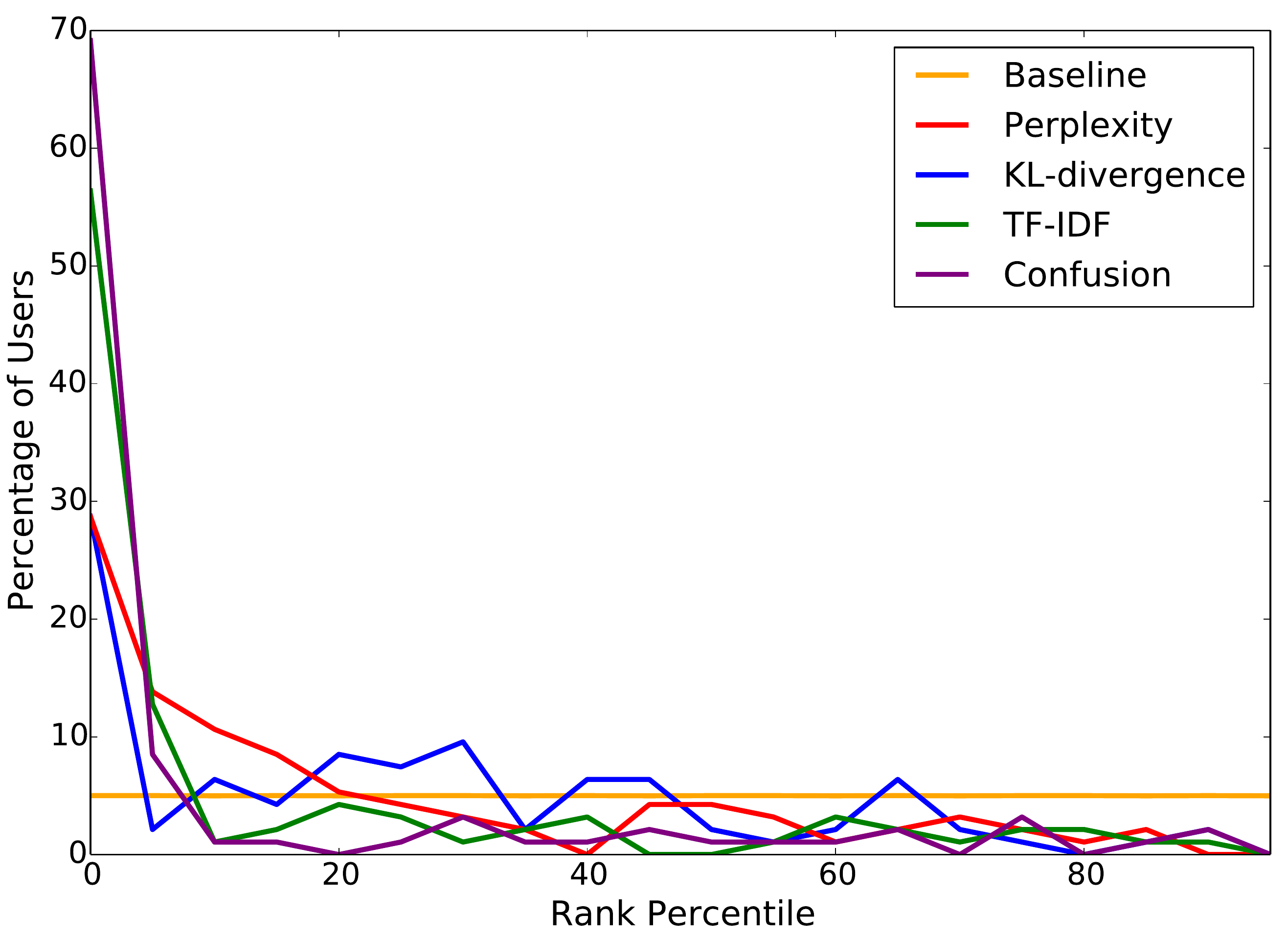}   
\caption{Rank percentiles of different combined temporal-linguistic models.}
\label{fig:rank_order_3}
\end{figure}

Another way to evaluate the performance of the different combined models is through the rank-statistics plot. This is shown in Figure \ref{fig:rank_order_3}. The figure shows the distribution of the ranks of the $5,612$ users for different combined models. The x-axis is the rank percentile (divided into bins of $5\%$), the y-axis is the percentage of the users that fall in each bin. For example, for the \emph{confusion model}, $69\%$ ($3880$) of the $5,612$ users were correctly linked between Twitter and Facebook when looking at the top $5\%$ ($281$) of the predictions by the model. From the figure, you can clearly see that the \emph{confusion model} is superior to the other models, with \emph{TF-IDF} a close second. You can also see from the figure that the rank plot for the random baseline is a horizontal line, with each rank percentile bin having $5\%$ of the users ($5\%$ because the rank percentiles were divided into bins of $5\%$).

\begin{table}

\centering
\caption{Performance of different combined models, tested on 5,612 users (11,224 accounts), sorted by accuracy. Best results are shown bold.}
\begin{tabular}{l|ll}
\hline
				& \multicolumn{2}{c}{Performance} \\
{Model}	 &{Accuracy} 		& {AverageRank}			\\ \hline
Random Baseline & 0.0002 & 2,806\\
KL-divergence & 0.11 	&1,741		\\
Perplexity & 0.11   & 1,303 	\\
TF-IDF	& 0.26	&	902  \\
Confusion	& \bf{0.31} 	&	\bf{745}\\
\hline
\end{tabular}

\label{table:comb_results}
\end{table}

\section{Evaluation Against Humans}
Matching profiles across social networks is a hard task for humans. It is a task on par with detecting plagiarism, something a non-trained person (or sometimes even a trained person) cannot easily accomplish. (Hence the need for the development of the field of stylometry in early Renaissance.) Be that as it may, we wanted to evaluate our model against humans to make sure that it is indeed outperforming them.

We designed an experiment to compare the performance of human judges to our best model, the temporal-linguistic confusion model. The task had to be simple enough so that human judges could attempt it with ease. For example, it would have been ludicrous to ask the judges to sort $11,224$ accounts into $5,612$ matching pairs. 

Thus, we randomly selected $100$ accounts from distinct users from our collection of $11,224$ accounts.  A unique list of $10$ candidate accounts was created for each of the $100$ accounts. Each list contained the correct matching account mixed in with $9$ other randomly selected accounts. The judges were then presented with the $100$ accounts one at a time and asked to pick the correct matching account from the list of $10$ candidate accounts. For simplicity, we did not ask the judges to do any ranking other than picking the one account that they thought matched the original account. We then measured the accuracy of the judges based on how many of the $100$ accounts they correctly matched. We had our model do the exact same task with the same dataset. A random baseline model would have a one in ten chance of getting the correct answer, giving it an accuracy of $0.10$.

We had a total of $3$ English speaking human judges from Amazon Mechanical Turk (which is an tool for crowd-sourcing of human annotation tasks) \footnote{https://www.mturk.com/}. For each task, the judges were shown the link to one of the $100$ account, and its $10$ corresponding candidate account links. The judges were allowed to explore each of the accounts as much as they wanted to make their decision (since all these accounts were public, there were no privacy concerns).
 
Table \ref{tab:human_eval} shows the performance of each of the three human judges, our model and the random baseline. Since the task is so much simpler than pairing $11,224$ accounts, our combined confusion model had a much greater accuracy than reported in the last section. With an accuracy of $0.86$, our model vastly outperformed even the best human judge, at $0.69$. Overall, our model beat the average human performance by $0.26$ ($0.86$ to $0.60$ respectively) which is a $43\%$ relative (and $26\%$ absolute) improvement.

 \begin{table}

\centering
\caption{Performance of the three human judges and our best model, the temporal-linguistic confusion model.}
\begin{tabular}{l|l}
\hline			
{Model}	 &{Accuracy} 					\\ \hline
Random Baseline & 0.10 \\
Human C	& 0.49  \\
Human B & 0.63    	\\
Human A & 0.69 		\\
Average Human & 0.60 \\
Combined Confusion & \bf{0.86} \\
\hline
\end{tabular}

\label{tab:human_eval}
\end{table}

\section{Discussion and Conclusions}
Motivated by the growing interest in matching user account across different social media and networking sites, in this paper we presented models for \emph{Digital Stylometry}, which is a method for matching users through stylometry inspired techniques. We used temporal and linguistic patterns of users to do the matching. 

We experimented with linguistic, temporal, and combined temporal-linguistic models using standard and novel techniques. The methods based on our novel \emph{confusion model} outperformed the more standard ones in all cases. We showed that both temporal and linguistic information are useful for matching users, with the best temporal model performing with an accuracy of $.10$ and the best linguistic model performing with an accuracy of $0.27$. Even though the linguistic models vastly outperformed the temporal models, when combined the temporal-linguistic models outperformed both with an accuracy of $0.31$. The improvement in the performance of the combined models suggests that although temporal information is dwarfed by linguistic information, in terms of its contribution to digital stylometry, it nonetheless provides non-overlapping information with the linguistic data.

Our models were evaluated on $5,612$ users with a total of $11,224$ accounts on Twitter and Facebook combined. In contrast to other works in this area, we did not use any profile information in our matching models. The only information that was used in our models were the time and the linguistic content of posts by the users. This is in accordance with traditional stylometry techniques (since people could lie or misstate this information). Also, we wanted to show that there are implicit clues about the identity of users in the content (language) and context (time) of the users' interactions with social networks that can be used to link their accounts across different services.

In addition to the technical contributions (such as our \emph{confusion model}), we hope that this paper is able to shed light on the relative ease with which seemingly innocuous information can be used to track users across social networks, even when signing up on different services using completely different account and profile information. In the future, we hope to extend this work to other social network sites, and to incorporate more sophisticated techniques, such as topic modelling and opinion mining, into our models.




\subsubsection*{Acknowledgments}
We would like to thank Ivan Sysoev for his help with developing the confusion model. We would also like to thank William Powers for sharing his insights on user privacy in the age of social networks. Finally, thanks to all the human annotators for their help with the evaluation of our models.

\bibliographystyle{splncs03}
\bibliography{socinfo}
\end{document}